\documentclass[conference]{IEEEtran}
\usepackage{amsmath}
\usepackage{amssymb}
\usepackage{footnote}

\begin{document}
\title{On Match Lengths and the Asymptotic Behavior of Sliding Window Lempel-Ziv Algorithm for Zero Entropy Sequences}
\author{\IEEEauthorblockN{Siddharth Jain}
\IEEEauthorblockA{Department of Electrical Engineering\\
Indian Institute of Technology Kanpur\\
Kanpur, India 208016\\
Email: sidjain@iitk.ac.in}
\and
\IEEEauthorblockN{Rakesh K. Bansal}
\IEEEauthorblockA{Department of Electrical Engineering\\
Indian Institute of Technology Kanpur\\
Kanpur, India 208016\\
Email: rkb@iitk.ac.in}}
\maketitle
\begin{abstract}
The Sliding Window Lempel-Ziv (SWLZ) algorithm has been studied from various perspectives in information theory literature. In this paper, we provide a general law which defines the asymptotics of match length for stationary and ergodic zero entropy processes. Moreover, we use this law to choose the match length $L_o$ in the almost sure optimality proof of Fixed Shift Variant of Lempel-Ziv (FSLZ) and SWLZ algorithms given in literature. First, through an example of stationary and ergodic processes generated by irrational rotation we establish that for a window size of $n_w$ a compression ratio given by $O(\frac{\log n_w}{{n_w}^a})$ where $a$ is arbitrarily close to $1$ and $0 < a < 1$, is obtained under the application of FSLZ and SWLZ algorithms. Further, we give a general expression for the compression ratio for a class of stationary and totally ergodic processes with zero entropy.
\end{abstract}
\section{Introduction}
The analysis of the string matching universal data compression algorithm in probabilistic setting was initiated by Wyner and Ziv~\cite{Ziv89} using the results from ergodic theory. This work was followed by the results on return times for stationary and ergodic sources by Ornstein and Weiss~\cite{Ornstein93}. Further, connections between return times and data compression have been obtained by Kontoyiannis~\cite{K}. Kim and Park~\cite{Kim} have given results on return time for \emph{zero entropy} ergodic processes generated by irrational rotation.

The Sliding Window Lempel Ziv (SWLZ) algorithm, which is very similar to the original string matching algorithm, was introduced by Wyner and Ziv~\cite{Ziv94} and its asymptotic optimality for the class of stationary and ergodic sources was proved in expected sense. Shields~\cite{Shields99} proved the  optimality of SWLZ algorithm for individual sequences by comparing the number of phrases of the SWLZ algorithm with that of the incremental parsing algorithm LZ-78~\cite{Ziv78}. On similar lines as Wyner and Ziv~\cite{Ziv94}, Jacob and Bansal~\cite{Bansal} proved the asymptotic optimality of the SWLZ algorithm in almost sure sense. However, it is true that the match length $L_o$ used by Wyner and Ziv and Jacob and Bansal can be used to prove the almost sure optimality of SWLZ algorithm for zero entropy processes but in this paper motivated by the behavior of SWLZ on periodic sequences (which is discussed in detail in section II) and recent results by Kim and Park~\cite{Kim} on recurrence properties for irrational rotations and in general for almost every interval exchange map with zero entropy~\cite{Kim1}, we show that for a general zero entropy process as well, it is possible to choose a larger match length $L_o$ in the proof of optimality of SWLZ algorithm, through which we establish results on convergence rate of the compression ratio of the SWLZ algorithm that are achievable for zero entropy sequences.

Lastras-Monta$\tilde{n}$o~\cite{Montano2006} showed that for a positive entropy finite order aperiodic and irreducible Markov source, SWLZ algorithm achieves a compression ratio of the order of $H+O(\sqrt{\frac{\log\log n_w}{\log n_w}})$ for a window of size $n_w$. In this paper, we obtain a finer statement for zero entropy cases. This is discussed in detail in section III and IV.

The remainder of this paper is organized as follows. In Section III we provide results on return times and match lengths and establish a theorem on match length. In Section IV subsections A and B, we prove the almost sure optimality of FSLZ and SWLZ algorithms respectively by making a different choice of match length $L_o$ for zero entropy processes as compared to the one chosen by Wyner and Ziv in~\cite{Ziv94} or Jacob and Bansal in~\cite{Bansal}. Further, this is followed by providing tighter upper bounds on convergence rate of the compression ratio for a class of zero entropy processes for these algorithms. In Section V, we present our conclusion.

\section{Motivation}
Consider $x =010101...,$ and $Tx = 101010......,$  to be two $n$ long sequences. Let $P(x) = P(Tx) = 0.5$ be the probability of occurence of $x$ and $Tx$ respectively. So, $\{x,Tx\}$ define the $2$ realizations of a stationary ergodic process which are periodic and entropy rate of the process is zero.

If LZ~78 is applied on the sequence $x$ then the the number of phrases generated are given by $c(n) \sim n^{\frac{1}{2}}$ for an $n$ long segment of the sequence $x$, which gives the compression ratio to be $\frac{c(n)\log c(n)+c(n)}{n}  \sim \frac{n^{\frac{1}{2}}\log n +n^{\frac{1}{2}}}{n} \sim \frac{\log n}{n^{\frac{1}{2}}}.$ However, if SWLZ algorithm is applied on the sequence $x$, $1$ bit will be required to specify the start in the initial window of size $2$ and $\log n$ bits will be required to specify the length of the match, since in one iteration complete match will be found in the initial window itself. So, in this case the compression ratio will be $\frac{\log n}{n}$, which converges faster as compared to the compression ratio in the case of LZ-78 applied on sequence $x$. This motivates the study of the behavior of SWLZ algorithm on aperiodic sequences generated by an ergodic process of zero entropy rate.

\section{Return Times and Match Lengths}

Let the sequence $x= \{x_n\}_{-\infty}^\infty$ be an instance of an ergodic source $X= \{X_n\}_{-\infty}^\infty$ with finite alphabet $A$. Here, $x_k^j$ denotes the segment $x_k,x_{k+1},......,x_j$ for sequence $x$. For $n > 0$, define
$$ R_n(x) = \min\{l: l > 0, x_{1}^{n} = x_{-l+1}^{-l+n}\}. $$
Thus, $R_n(x)$ is the first return time of the word $x_{1}^{n}$ in the past.

$$ L_n(x) = \max\{j: j > 0, x_1^j = x_{-k+1}^{-k+j}, k = 1,2,....,n\}.$$
Thus $L_n(x)$ is the length of the longest possible match in the previous $n$ symbols $x_{-n+1}^0$.
On the asymptotic behavior of $R_n(x)$ and $L_n(x)$ we have the following,\\
\textbf{Theorem 1 (Ornstein and Weiss):}
With probability 1, for a  stationary and ergodic source $X$
 $$\lim_{n \rightarrow \infty} \frac{\log R_n(x)}{n} = H(X); \lim_{n \rightarrow \infty} \frac{\log n}{ L_n(x)} = H(X). $$
where $H(X)$ is the entropy of the source X.\\
As a general case, consider a stationary and ergodic process for which the asymptotics of the first return time follow
\begin{equation}\label{101}
\lim_{n \rightarrow \infty} \frac{\log R_n(x)}{f(n)} = c.
\end{equation}
where $f(n)\leq n$ is an invertible and increasing function of $n$ and $c > 0$ is a constant. It is notable here that for a zero entropy process $f(n) = o(n)$.\\
\textbf{Theorem 2:} With probability 1 for a stationary and ergodic process satisfying Eq. (\ref{101})
\begin{equation*}
\lim_{n \rightarrow \infty} \frac{\log n}{f(L_n(x))} = c.
\end{equation*}
\textbf{Proof of Theorem 2:}
Let $n$ and $m$ be natural numbers such that
$n = L_m(x)$. Hence, by definition of a match length we have
\begin{equation}\label{Inequality}
R_{n+1}(x) > m ~and~ R_n(x) \leq m.
\end{equation}

Therefore, using Inequality (\ref{Inequality}) we have
\begin{equation}\label{Sandwich}
\frac{\log R_{n+1}(x)}{f(n)} > \frac{\log m}{f(L_m(x))} \geq \frac{\log R_n(x)}{f(n)}.
\end{equation}
Now we apply the limit $ n \rightarrow \infty$ on inequality (\ref{Sandwich}) and hence using Sandwich theorem and making a change in the name of variable from $m$ to $n$ we have obtained the statement of Theorem 2.

Now we shift our attention to zero entropy sources. For this purpose, we first consider the sources generated by an irrational rotation.

Let $v ~\epsilon ~\mathcal{R}$ and let $\|v\|$ denote its distance to the nearest integer, i.e.,
$ \|v\| = \min_{n\epsilon Z} |v-n|. $

Let $\theta$ be an irrational number in $(0,1)$. Consider $T : [0,1) \rightarrow [0,1)$ to be an irrational rotation by $\theta$. i.e., $ T(x) = (x+\theta) ~mod ~1. $
$T$ preserves the Lebesgue measure on $X = [0,1)$.
For an irrational $\theta ~ \epsilon ~(0,1)$, define
$ \eta(\theta) = \sup \{ t>0 : \liminf_{i\rightarrow\infty} i^t\|i\theta\| = 0\}.$
In general, $\eta(\theta) \geq 1$ and $= 1$ for almost every $\theta$~\cite{Kim}. For generating a sequence $x_1^n$ by irrational rotation a partition $\mathcal{P} = \{E, E^c\}$ is defined on $[0,1]$, such that $x_i = 0$ if $T^{i-1}x ~\epsilon~ E$ and 1 otherwise. \\
\textbf{Theorem 3 (Kim and Park):}
For almost every sequence generated by rotation by an irrational $\theta ~\epsilon ~(0,1)$:
\begin{equation}\label{1}
\liminf_{n\rightarrow\infty} \frac{\log R_n(x)}{\log n} = \frac{1}{\eta(\theta)};~
\limsup_{n\rightarrow\infty} \frac{\log R_n(x)}{\log n} = 1.
\end{equation}
which implies that if $\theta$ is such that $\eta(\theta) = 1$ then the limit exists and is given by:
\begin{equation}\label{3}
\lim_{n\rightarrow\infty} \frac{\log R_n(x)}{\log n} = 1.
\end{equation}
Now, applying Theorem 2 by choosing the function $f(n) = \log n$, we immediately deduce the following corollary\\
\textbf{Corollary 1:} For a stationary and ergodic process generated by irrational rotation where $\eta(\theta) = 1$  with probability 1, $ \lim_{n \rightarrow \infty} \frac{\log n}{\log L_n(x)} = 1. $\\
\textbf{Critical Observation 1:} Consider a stationary and ergodic process satisfying the law given by Eq. (\ref{101})

By Eq. (\ref{101}) and Egoroff's Theorem~\cite{Royden68}, for a given $\delta > 0$ and $\epsilon > 0$, $\exists$ a set $F_1$ of sequences $x$ with $P(F_1) > 1- \delta$ and an integer $N_o(\epsilon, F_1)$ s.t.
\begin{equation*}
\begin{split}
0 < \frac{\log R_n(x)}{f(n)} < c+ \epsilon  ~~~~\forall ~~n \geq N_o(\epsilon, F_1), ~x~\epsilon~F_1\\
\Rightarrow \frac{\log R_n(x)}{c+\epsilon} < f(n)
          \Rightarrow R_n(x) < 2^{(c+\epsilon)f(n)}.
\end{split}
\end{equation*}
Now, if $n_w$ is the window size and $n_w = 2^{(c+\epsilon)f(L_o)}$, then we have $R_{L_o}(x) < n_w$ which implies that a $\emph{match}$ of length $L_o$ is bound to be found in $n_w$ symbols and hence the match length is given by
\begin{equation}\label{102}
L_o = \lfloor f^{-1} (\frac{\log n_w}{c+\epsilon})\rfloor.
\end{equation}
For example in a positive entropy case, where $f(n) = n$ and $c = H$, $L_o =\lfloor{\frac{\log n_w}{H+\epsilon}}\rfloor$ and  in case of irrational rotation where as given by Eq. (\ref{1}) in Theorem 3, $f(n) = \log n$, $ c = 1$ and $\eta(\theta) = 1$, we have
\begin{equation}\label{7}
L_o = \lfloor{n_w}^{\frac{1}{1+\epsilon}}\rfloor.
\end{equation}
\section{Convergence Rate of the Compression Ratio}
It has been proved in~\cite{Montano2006} that the compression ratio given by $H+ O(\sqrt{\frac{\log{\log{n_w}}}{\log n_w}})$ is achieved by SWLZ algorithm for an aperiodic and irreducible Markov source with positive entropy.
Following the discussion in Section III, in this section we prove that a compression ratio given by $O(\frac{\log n_w}{{n_w}^a})$, where $ 0 < a < 1$, is achieved by Fixed Shift Variant of SWLZ (FSLZ) algorithm~\cite{Bansal} and SWLZ algorithm for stationary and ergodic processes generated by an irrational rotation. Further, we give a general expression of the compression ratio for zero entropy cases under the setting described by Eq. (\ref{101}) for both FSLZ and SWLZ algorithms.

In the remainder of this section, we illustrate through the example of stationary and ergodic processes generated by irrational rotations that a compression ratio given by $O(\frac{\log n_w}{{n_w}^a})$ where $ 0 < a < 1$ and $a$ is arbitrarily close to $1$, is achieved by FSLZ and SWLZ algorithms respectively, following the proofs of optimality of these algorithms as given in~\cite{Ziv94} and \cite{Bansal} but choosing $L_o$ given by Eq. (\ref{7}). It is emphasized here that the $L_o$ chosen is in contrast to what is chosen in \cite{Ziv94} and \cite{Bansal}. Moreover, following this illustration, a general expression for the compression ratio for a class of zero entropy stationary and totally ergodic processes is given under a certain restriction on the convergence rate of the law given in Eq. (\ref{101}).

\subsection{Fixed Shift Variant of SWLZ Algorithm (FSLZ)}
We first consider FSLZ algorithm introduced in~\cite{Bansal} to gain insight into the working of the universal SWLZ algorithm. Let us consider the string $x_1^N$ consisting of the first $N$ symbols of the sequence. Following the discussion in Section III. Using critical observation 1, let us take for $\epsilon > 0$, the match length given by Eq. (\ref{7}).
Note the contrast in the choice of match length that is used in~\cite{Ziv94} and~\cite{Bansal} for positive entropy stationary and ergodic processes.The algorithm is described as follows:
\begin{enumerate}
  \item  \emph{Initialization}: A window of size $n_w$ is fixed. Let $j =1$ and $n_j = n_w$. The first $n_w$ symbols are transmitted without compression. Let $x_1^{n_1}$ be the current window.
  \item  \emph{Matching}: At the $j$th step if $n_j+L_o > N$, terminate the algorithm, else if $x_{n_j+1}^{n_j+L_o} = x_{n_j-n_w+k}^{n_j-n_w+k+L_o-1}$ for any $k ~\epsilon ~{1,2,....,n_w}$ then there is a match, i.e., sequence $x_{n_j+1}^{n_j+L_o}$ has a match in the current window.
  \item \emph{Coding}: If there is a match, let $s_j$ be the location of the match in the current window. Then $\lceil \log n_w\rceil$ bits are required to code $s_j$. If a match is not found the number of bits required is $\beta L_o$. In addition, a one bit flag is used to specify if the match is found or not found.
  \item \emph{Sliding}: $n_{j+1}= n_{j}+L_o$, i.e., the current window now becomes $x_{n_{j+1}-n_w+1}^{n_{j+1}}$.
\end{enumerate}
Repeat the steps 2, 3 and 4 until the sequence $x_1^N$ terminates at the $(m+1)$-th step. The remaining $N-n_w-mL_o <L_o$ symbols are encoded without compression. Thus, a total of $N-mL_o < n_w+L_o$ are encoded without any compression. Let $m_1$ be the number of blocks that have a match and $m_2$ be the number of blocks that do not have a match. One bit per block is used to denote if it is good or bad. Hence, taking overheads into account, the number of bits per symbol required by the code is given by
\begin{equation}\label{8}
\begin{split}
R_{FSLZ} &= \frac{1}{N}\Big[(N - mL_o)\beta + m_1\lceil \log n_w \rceil + m_2\beta L_o +m\Big]\\&
< \frac{1}{N}\Big[(n_w + L_o)\beta + m_1\lceil \log n_w \rceil + m_2\beta L_o + m\Big].
\end{split}
\end{equation}
The first term in the expression converges to 0 as $N \rightarrow \infty$.
Also, we have
\begin{equation}\label{9}
m_1 \leq m < \frac{N}{L_o}.
\end{equation}
Using Eq. (\ref{7}) and (\ref{9}) we have,
\begin{equation}\label{10}
\frac{m_1\lceil \log n_w\rceil}{N} < \frac{\lceil \log n_w\rceil}{L_o} < \frac{\lceil \log n_w\rceil}{{n_w}^{\frac{1}{1+\epsilon}}-1}.
\end{equation}
which converges to 0 as $n_w \rightarrow \infty$.
Therefore, using Eq. (\ref{10}) we have
\begin{equation}\label{11}
  \lim_{n_w \rightarrow \infty} \frac{m_1\lceil \log n_w\rceil}{N} = 0.
\end{equation}
 Hence, we have the following\\
\textbf{Critical Observation 2: }The second term in the expression of $R_{FSLZ}$ (Eq. (\ref{8})) converges to $0$ for all $\epsilon > 0 $ as $n_w\rightarrow\infty$ and the convergence rate is given by $O(\frac{\log n_w}{n_w^{\frac{1}{1+\epsilon}}})$.

Now , define $ G = \{ x: R_{L_o}(x) > n_w\}.$ Thus, if $ x~\epsilon~ G$, then $x$ does not have a match in the previous $n_w$ symbols. We use $1_G$ to denote the indicator function of $G$. Since $T$ is \emph{totally} ergodic (i.e., $T^k$ is ergodic for all k as $k\theta$ is irrational for all $k$), Z the \emph{shift transformation} is also totally ergodic. By the Ergodic Theorem~\cite{Shields96} we have (almost surely)
\begin{align}\label{12}
\frac{m_2L_o}{N} < \frac{m_2}{m}
                 &= \frac{\sum_{j=0}^{m-1} 1_G(Z^{jL_o+n_w}(x))}{m}
                  \rightarrow \mu(G).
\end{align}
Now,
\begin{equation*}
\begin{split}
G &= \{ x: R_{L_o}(x) > n_w \} \\
  &\subset \{ x: \frac{\log R_{L_o}(x)}{\log L_o} > 1+\epsilon \}. \\
\end{split}
\end{equation*}
Hence, by Theorem 3 $\lim_{n_w \rightarrow \infty} \mu(G) = 0$.\\
\textbf{Critical Observation 3: }It is established in~\cite[Sec. 2 page-3944, Sec. 3 page-3948, 3949]{Kim} that the convergence rate of the probability of the set $G$ is atleast as fast as $\frac{1}{{n_w}^b}$, where $b > a$, for $a$  defined in result 1 and result 3 below.

Combining (\ref{8}), (\ref{9}), (\ref{11}), (\ref{12}) and $\mu(G) \rightarrow 0$, we have for almost every sequence generated by an irrational rotation, the FSLZ algorithm is asymptotically optimal. i.e.,
$\lim_{n_w \rightarrow \infty} \lim_{N\rightarrow \infty} R_{FSLZ} = 0$.

Using critical observations 2 and 3, we can now state the following\\
\textbf{Result 1:} For a stationary and ergodic process generated by an irrational rotation a compression ratio given by $O(\frac{\log n_w}{{n_w}^a})$ is achieved by FSLZ algorithm, where $a = \frac{1}{1+\epsilon}$ and $\epsilon > 0$ is arbitrary.

Now, we consider a general setting where the process is assumed to be stationary and totally ergodic with zero entropy and the asymptotics of the first return time follow the law given by Eq. (\ref{101}). In such a case following the proof of FSLZ algorithm considered above, we choose $L_o$ given by Eq. (\ref{102}). Imitating the proof given above with this choice of $L_o$ yields us a term of the form $ \frac{\lceil \log n_w \rceil}{f^{-1}(\frac{\log n_w}{c+\epsilon})-1}$ and the set $G \subset \{x:\frac{\log R_{L_o}(x)}{f(L_o)} > c + \epsilon\}.$
Since $f(n) = o(n)$ , we have
\begin{equation}\label{103}
\lim_{n_w \rightarrow \infty} \frac{\lceil\log n_w\rceil}{f^{-1}(\frac{\log n_w}{c+\epsilon})-1} = 0.
\end{equation}
Also, using Eq. (\ref{101}), $\lim_{ n_w \rightarrow \infty} \mu(G) = 0.$\\
If we further assume\\
\textbf{Assumption 1: }The rate of convergence of $\mu(G)$ to $0$, is at least as fast as $ \frac{\lceil \log n_w\rceil}{\lfloor f^{-1}(\frac{\log n_w}{c+\epsilon})\rfloor}$, then we have\\
\textbf{Result 2:} For a stationary and totally ergodic process for which the asymptotics of first return time follow the law given by Eq. (\ref{101}) and assumption 1 holds, a compression ratio given by $ O(\frac{\log n_w}{f^{-1}(\frac{\log n_w}{c+\epsilon})})$ is achieved by FSLZ algorithm. Here $\epsilon > 0$ is arbitrary.

For example, if $f(n) =\sqrt{n}$ the compression ratio for FSLZ algorithm is given by $O(\frac{{(c+\epsilon)}^2}{\log n_w}).$

\subsection{The Sliding Window Lempel-Ziv Algorithm}
Consider a sequence $x = \{x_{n}\}_1^\infty$ which is sequentially made available to the encoder. Let $A$ be the finite set of alphabet for the sequence $x$. Let $\beta \triangleq \lceil\log|A|\rceil$. If $S$ is a finite set then $|S|$ denotes the cardinality of S.  Sliding Window LZ is now described ~\cite{Ziv94} \cite{Bansal} :

Consider the string $x_1^N$ consisting of the first $N$ symbols of the sequence $x = \{x_{n}\}_1^\infty$. The algorithm works as follows:
\begin{enumerate}
  \item  \emph{Initialization}: A window of size $n_w$ is fixed. Let $j =1$ and $n_j = n_w$. The first $n_w$ symbols are transmitted without compression. Let $x_1^{n_1}$ be the current window.
  \item  \emph{Matching}: At the $j$th step let $L_j$ be the largest integer such that the copy of $x_{n_j+1}^{n_w+L_j}$ begins in the current window and $n_j+L_j\leq N$. Let $s_j$ denote the starting index of the match in the current window. The matched phrase $x_{n_j+1}^{n_w+L_j}$ is denoted by $x^{(j)}$. If a match is not found $ L_j =1$, $s_j = 0$.
  \item \emph{Coding}: If $s_j >0$, the length $L_j$ of the matched code can be specified using $\gamma\log (L_j+1)$ bits using the integer code described in ~\cite{Ziv94}. The matched location $s_j$ can be specified using $\lceil\log n_w\rceil$ bits. If $s_j =0$ or if, using the above procedure, the total number of bits needed to represent a phrase exceeds $\beta L_j$ then $\beta L_j$ bits are used for encoding. A one bit flag is used to denote which of these two encoding schemes was used.
  \item \emph{Sliding}: For the next window $n_{j+1} = n_j + L_j$ and the window for the next iteration is $x_{n_{j+1}-n_w+1}^{n_{j+1}}$.

\end{enumerate}
Repeat the steps 2,3 and 4 until the sequence $x_1^N$ is exhausted.

Let $B(x^{(j)}) + 1$ denote the number of bits required to encode the $j$-th phrase. Then, we have
\begin{equation}\label{5}
B(x^{(j)}) = \min{\{\gamma\log (L_j+1) + \lceil\log n_w\rceil, \beta L_j \}}.
\end{equation}
If the total number of phrases is $c(N)$ then, the number of bits required per symbol is:
\begin{equation}\label{6}
R_{SWLZ}= \frac{1}{N}\Big[n_w\beta +\sum_{j=1}^{c(N)} B(x^{(j)})+c(N)\Big].
\end{equation}
Here, we analyze the performance of SWLZ algorithm on zero entropy sequences generated by irrational rotation Transformation T. The idea behind the analysis is inspired from the method of partitioning of sequences used by~\cite{Ziv94}. However, unlike them the class of partitions considered here are obtained by shifting the partition used by them. By this the optimality result is obtained in almost sure sense~\cite{Bansal}. Consider the intervals defined as $ I_r = [r+n_w,r+n_w+L_o-1],~ r = 1,2,...,N' $. Here, $ N' = N-L_o-n_w +1.$ Interval $I_r$ is bad if a copy of ${(x_i)}_{i \epsilon I_r}$ does not begin in the string $ n_w -L_o $ symbols preceeding it. Let $m$ be the number of such bad intervals.
Now, we define
$ G = \{x:R_{L_o}(x) > n_w - L_o\}.$
So , if $ x ~\epsilon~ G $, then $[1,L_o]$ is a bad interval.
By Ergodic Theorem almost surely,
\begin{equation}\label{15}
\frac{m}{N} < \frac{m}{N'} = \frac{\sum_{j=0}^{N'-1}1_G(Z^j(x))}{N'} \rightarrow \mu (G). ~~ (as ~N \rightarrow \infty)
\end{equation}
Now, $\quad \quad ~~~G = \{ x: R_{L_o}(x) > n_w - L_o \}$\\
\begin{equation*}
 \begin{split}
  &\subset \{ x: R_{L_o}(x) > L_o^{1+\frac{\epsilon}{2}}  \}\\& = \{ x: \frac{\log R_{L_o}(x)}{\log L_o} > 1+ \frac{\epsilon}{2} \}.
\end{split}
\end{equation*}
Hence by Theorem 3,
\begin{equation}\label{18}
\lim_{n_w \rightarrow \infty} \mu(G) = 0.
\end{equation}
which implies from (\ref{15})
$\lim_{n_w \rightarrow \infty} \frac{m}{N} = 0.$
As, already specified in critical observation 3, we have the convergence of rate of $\mu(G)$ to $0$ (as $n_w \rightarrow \infty$) is at least as fast as $\frac{1}{{n_w}^b},$ where $b > a$.

Let us consider the SWLZ parsing of the sequence $x_1^N$. A phrase is defined to be internal with respect to interval $I_r$ if it begins and ends in the interval $[r+n_w, r+ L_o+n_w-2]$~\cite{Ziv94}. Moreover, an interval $I_r$ with an internal phrase is bad~\cite{Ziv94}.

For $k = 0,1,2.....,L_o - 1$ , a partition $ \mathcal{P}_k$ is defined as given in~\cite[Section V]{Bansal}.
Let the collection of internal phrases corresponding to $\mathcal{P}_k$ together with initial and final segment be denoted by set $S_k$ and let $m_k$ be the number of bad intervals. Then, using bound on $\sum_{j \epsilon S_k} L_j$ and defining $m_{k_o} = \min_k \{m_k\}$ as in \cite[Section V]{Bansal}, $L_om_{k_o} \leq m$ is obtained.\\
Using Eq. (\ref{6}), the number of bits per symbol for SWLZ algorithm is given by
$$R_{SWLZ}= \frac{1}{N} \Big[n_w\beta +\sum_{j \epsilon S_{k_o}} (\beta L_j+1) + \sum_{j \epsilon S_{k_o}^c} (B(x^{(j)})+1)\Big]$$
\begin{equation}\label{24}
\begin{split}
\Rightarrow R_{SWLZ} &\leq \frac{1}{N} \Big[(n_w + 2 L_o)(\beta + 1) +(\beta+1) m \\&
  +\sum_{j \epsilon S_{k_o}^c} B(x^{(j)})+|S_{k_o}^c|\Big].
\end{split}
\end{equation}
 The terms $\frac{n_w(\beta + 1)}{N}$ and $\frac{2(\beta+1) L_o}{N}$ converge to 0 as $N \rightarrow \infty$. From Eq. (\ref{18}) it is evident that the second term in the expression for $R_{SWLZ}$ given by Eq. (\ref{24}) converges to 0 (as $ N \rightarrow \infty $ followed by $n_w \rightarrow \infty$). Now, we consider the last term in the expression for $R_{SWLZ}$ given in Eq. (\ref{24}) i.e., ~$\frac{1}{N} \sum_{j \epsilon S_{k_o}^c} B(x^{(j)}).$
Using the method given in~\cite{Ziv94}, let $ d =|S_{k_o}^c| $.
Since, $S_{k_o}^c$ comprises of non internal phrases corresponding to partition $\mathcal{P}_{k_o}$ together with the initial and final segments,the non-internal phrase should end at the last index of any interval that belongs to $\mathcal{P}_{k_o}$.
\begin{equation}\label{27}
|S_{k_o}^c| = d \leq \frac {N'}{L_o} < \frac{N}{L_o}.
\end{equation}
Using Eq. (\ref{27}) we have $\lim_{n_w \rightarrow \infty} \frac{|{S_{k_o}}^c|}{N} = 0 $. Further, imitating the method used in~\cite{Ziv94}, we have
\begin{equation}\label{28}
\begin{split}
\frac{1}{N} \sum_{j \epsilon S_{k_o}^c} B(x^{(j)}) &= \frac{1}{N}\sum_{j \epsilon S_{k_o}^c} \Big\{\gamma \log(L_j + 1) + \lceil \log n_w \rceil\Big\}  \\      &= \frac{1}{N}|S_{k_o}^c| \lceil \log n_w \rceil  + \frac{1}{N}\sum_{j \epsilon S_{k_o}^c} \gamma \log(L_j + 1)  \\
                                    &\leq \frac{1}{L_o} \lceil \log n_w \rceil + \frac{d\gamma}{N} \sum_{j \epsilon S_{k_o}^c} \frac{1}{d}\log(L_j + 1) ~(a_1)\\
                                      &\leq \frac{\lceil \log n_w \rceil}{L_o} + \frac{d\gamma}{N} \log(\frac{1}{d} \sum_{j \epsilon S_{k_o}^c} (L_j + 1)) ~(a_2)\\
                                       &\leq \frac{\lceil \log n_w \rceil}{n_w^{\frac{1}{1+\epsilon}}-1} + \frac{\gamma}{L_o} \log(L_o + 1) ~(a_1)\\
                                     &= 0. ~~(as ~n_w \rightarrow \infty)
\end{split}
\end{equation}
Here, $(a_1)$ means using Eq. (\ref{27}) and $(a_2)$ means using the concavity of log function.\\
\textbf{Critical Observation 4: } It is evident in Eq. (\ref{28}) that the compression ratio converges to 0 for good phrases at the rate given by $O(\frac{\log n_w}{n_w^{\frac{1}{1+\epsilon}}})$, for every $\epsilon > 0.$
Hence, combining results of Eq. (\ref{18}) and (\ref{28}) we get for almost every sequence generated by an irrational rotation, the SWLZ algorithm is asymptotically optimal i.e.,
$\lim_{n_w \rightarrow \infty} \lim_{N \rightarrow \infty} R_{SWLZ}(x) = 0 ~a.s.$\\
So using critical observations 3 and 4 we have\\
\textbf{Result 3:} For stationary and ergodic processes generated by an irrational rotation a compression ratio given by $O(\frac{\log n_w}{{n_w}^a})$ is achieved by SWLZ algorithm, where $a = \frac{1}{1+\epsilon}$ and $\epsilon > 0$ is arbitrary.

Now, we consider a general setting where the process is assumed to be stationary and ergodic with zero entropy and the asymptotics of the first return time follow the law given by Eq. (\ref{101}). In such a case following the proof of SWLZ considered above, we choose $L_o$ given by Eq. (\ref{102}). Imitating the proof given above with this choice of $L_o$ yields us terms of the form $\frac{\lceil \log n_w \rceil}{f^{-1}(\frac{\log n_w}{c+\epsilon})-1} + \frac{\gamma}{L_o} \log(L_o + 1)$ and the set $$G \subset \{x:\frac{\log R_{L_o}(x)}{f(L_o)} > c + \frac{\epsilon}{2}\}.$$ Since $f(n) = o(n)$
\begin{equation}\label{104}
\lim_{n_w \rightarrow \infty}\frac{\lceil \log n_w \rceil}{f^{-1}(\frac{\log n_w}{c+\epsilon})-1} + \frac{\gamma}{L_o} \log(L_o + 1) = 0.
\end{equation}
If $f(n) \geq \log n$, then $L_o < n_w$ for every $\epsilon > 0$,
the compression ratio is given by $O(\frac {\log n_w}{f^{-1}(\frac{\log n_w}{c+\epsilon})})$ for every $\epsilon > 0$. Also we have using Eq. (\ref{101}) $\lim_{n_w \rightarrow \infty} \mu(G) = 0.$ Hence, we have the following\\
\textbf{Result 4:} For a stationary and totally ergodic process with zero entropy such that the asymptotics of its first return time follow the law given by Eq. (\ref{101}) with $f(n) \geq \log n$ and assumption 1 holds a compression ratio given by $O(\frac{\log n_w}{f^{-1}(\frac{\log n_w}{c+\epsilon})})$ is achieved by SWLZ algorithm.
For example, if $f(n) = \sqrt[3]{n}$, the compression ratio is given by $O(\frac{{(c+\epsilon)}^3}{{\log n_w}^2}).$ Here $\epsilon > 0$ is arbitrary.
\section{Conclusion}
In this paper, we state Theorem 2 on match length which generalizes the match length result given by Ornstein and Weiss~\cite{Ornstein93} for stationary and ergodic processes. Next, for the class of zero entropy processes  we establish through corollary 1, the behavior of match length asymptotics for irrational rotation and a general zero entropy stationary and ergodic case. Further, we imitated the proofs given in~\cite{Ziv94} and~\cite{Bansal} of FSLZ and SWLZ algorithms by choosing a $L_o$ given by Eq. (\ref{102}) in contrast to \emph{their} choice and showed that these algorithms achieve faster convergence rate of the compression ratio for \emph{zero entropy} sequences as compared to those with positive entropy. It will be an interesting problem to look for totally ergodic processes that display the behavior given by Eq. (\ref{101}) where the function $f(n)$ is different from $\log n $. Also, it will be of interest to determine the class of irrational rotations and partitions $\mathcal{P}$ used to generate the sequences for which compression ratio converges to zero, uniformly.
\section*{Acknowledgement}
The authors wish to thank T. Jacob for giving his thoughts on match lengths for zero entropy processes.

\end{document}